\documentclass[letterpaper,pra,notitlepage,superscriptaddress, twocolumn, nofootinbib, showpacs, eqsecnum]{revtex4-1}

\usepackage {amsmath}
\usepackage{graphicx}
\usepackage{braket}
\usepackage{amsthm}
\usepackage{amssymb}
\usepackage{color}
\usepackage{mathtools}
\usepackage{tikz}
\usepackage{leftidx}
\usepackage{bbold}
\usetikzlibrary{decorations}
\usetikzlibrary{patterns}

\usepackage[colorlinks=true,urlcolor=blue, linkcolor=blue]{hyperref}

\providecommand{\abs}[1]{\lvert#1\rvert}

\def\({\left(}
\def\){\right)}

\def\fig#1{Fig.~\ref{fig:#1}}

\DeclareMathOperator{\tr}{tr}

\usepackage{bm}

\newcommand{\id}{I}
\newcommand{\1}{\mat{\id}}
\newcommand{\mat}[1]{\bm{\mathrm{#1}}}

\graphicspath{ {./Figures/} }



\def\Id{1\!\mathrm{l}}
\newcommand{\gateset}{\mathsf{G}}
\newcommand{\Clifford}{\mathsf{C}}
\newcommand{\Pauli}{\mathsf{P}}
\newcommand{\Clifsub}{\mathsf{T}}
\newcommand{\uC}{U}
\newcommand{\cC}{\mathcal{U}}
\newcommand{\cD}{\mathcal{D}}
\newcommand{\cM}{\mathcal{M}}
\newcommand{\cX}{\mathcal{X}}
\newcommand{\cZ}{\mathcal{Z}}
\newcommand{\cH}{\mathcal{H}}

\newcommand{\cA}{\mathcal{A}}
\newcommand{\cQ}{\mathcal{Q}}

\newcommand{\cI}{\mathcal{I}}

\DeclareMathOperator*{\comp}{\bigcirc}

\begin{document}

\title{Randomized benchmarking in measurement-based quantum computing}
\author{Rafael N. Alexander}
 \email[Email: ]{rafael.alexander@sydney.edu.au}
\affiliation{Centre of Engineered Quantum Systems, School of Physics, The University of Sydney, Sydney, NSW 2006, Australia}
\affiliation{School of Science, RMIT University, Melbourne, VIC 3001, Australia}
\author{Peter S. Turner}
\affiliation{School of Physics and Department of Electrical and Electronic Engineering, University of Bristol,\\ HH Wills Laboratory, Tyndall Avenue, Bristol BS8 1TL, United Kingdom}
\author{Stephen D. Bartlett}
\affiliation{Centre of Engineered Quantum Systems, School of Physics, The University of Sydney, Sydney, NSW 2006, Australia}
\date{11 August 2016}

\begin{abstract}
Randomized benchmarking is routinely used as an efficient method for characterizing the performance of sets of elementary logic gates in small quantum devices.
In the measurement-based model of quantum computation, logic gates are implemented via single-site measurements on a fixed universal resource state.
Here we adapt the randomized benchmarking protocol for a single qubit to a linear cluster state computation, which provides partial, yet efficient characterization of the noise associated with the target gate set.
Applying randomized benchmarking to measurement-based quantum computation exhibits an interesting interplay between the inherent randomness associated with logic gates in the measurement-based model and the random gate sequences used in benchmarking.
We consider two different approaches: the first makes use of the standard single-qubit Clifford group, while the second uses recently introduced (non-Clifford) measurement-based 2-designs, which harness inherent randomness to implement gate sequences.
\end{abstract}
\maketitle

\section{Introduction}
\label{sec:intro}
In the measurement-based model~\cite{Raussendorf2001}, quantum computation proceeds via adaptive single-site measurements on an entangled resource state of many qubits such as the \emph{cluster state}~\cite{Briegel2001}.  
The computational power of this model is equivalent to standard approaches to universal fault-tolerant quantum computation, assuming that all operations can be implemented with sufficiently small error~\cite{Aliferis2006, Dawson2006, Dawson2006a}. 
Because this model does not require an on-demand entangling gate, it is appealing for candidate physical architectures where such gates cannot be performed deterministically. 
The leading example is linear-optical quantum computing (LOQC)~\cite{Knill2001, Kok2007, Nielsen2004}, where the basic building blocks are single-photon sources, linear optics, and photon-number resolving detectors with feedforward.

As quantum devices with progressively smaller error rates are developed, there is a growing need for techniques to efficiently characterize the noise associated with elementary components such as logic gates. 
Although it may sound desirable, a complete description of the error processes of a quantum device is prohibitively expensive due to the exponentially bad scaling in size~\cite{Chuang1997, Poyatos1997}. 
An additional concern is how to observe gate errors in the presence of noise from state preparation and measurement (SPAM), which often dominate.  
The randomized benchmarking (RB) protocol~\cite{Knill2008, Magesan2011, Magesan2012} is a technique that allows for efficient, partial characterization of a target gate set while being insensitive to noise from SPAM~\cite{Wallman2014}. 


Randomized benchmarking performs well with realistic noise using only small data sets~\cite{Epstein2014, Granade2015, Wallman2014}.  The basic protocol has been extended to include tests for time dependence, non-Markovianity~\cite{Wallman2014,Epstein2014,Fogarty2015,Ball2015}, robustness to leakage errors~\cite{Chasseur2015}, reconstruction of the unital part of general completely positive trace-preserving (CPTP) maps~\cite{Kimmel2014}, and extracting tomographic data from quantum gates~\cite{Johnson2015}. 

Here we adapt the original RB protocol to the setting of measurement-based quantum computation (MBQC). 
By combining ideas from RB and MBQC on linear cluster states, we provide two protocols for estimating the average gate fidelity for two different single-qubit gate sets. 
The first gate set is the single-qubit Clifford group, and the second is the recently proposed measurement-based exact 2-design~\cite{Turner2016}, which leverages the intrinsic randomness of MBQC to implement random sequences of gates.
Our schemes fully inherit the advantages of the RB protocol, namely that 
the average gate fidelity can be computed efficiently and with low sensitivity to errors in preparation of the (logical) input and final measurement readout~\cite{Magesan2012}.  
 
The structure of the paper is as follows.  We review the RB protocol and MBQC in Sec.~\ref{sec:background}. 
We discuss our protocols for implementing RB on a linear cluster state with the Clifford group and with the measurement-based 2-designs in Sec.~\ref{sec:RBinMBQC}.
\section{Background}
\label{sec:background}
Here we review the standard RB protocol~\cite{Magesan2012}, and fix our notation.

\subsection{Preliminaries}
\label{sec:prelims}
Consider a $(d=2^{n})$-dimensional Hilbert space $(\mathbb{C}^2)^{\otimes n}$ corresponding to an $n$-qubit system.  A unitary operation (gate) is denoted by $U$; the corresponding superoperator that acts on density matrices $\rho$ is denoted by $\mathcal{U}(\rho) = U \rho U^{\dagger}$. We denote $\mathcal{U}^{\dagger}(\rho) = U^{\dagger} \rho U$ and $\mathcal{U}^{m}(\rho) = U^{m} \rho {U^{m}}^{\dagger}$.  General (non-unitary) superoperators are denoted $\mathcal{D}$, $\mathcal{E}$, etc., and in addition we use $\tilde{\mathcal{U}}$ to denote a noisy approximation to the ideal unitary gate $\mathcal{U}$.  
Common unitary gates we will see include the $X$, $Y$, and $Z$ Pauli matrices, the Hadamard gate $H$, the controlled-Z $(CZ)$ gate ($\ket{0}\!\!\bra{0}\otimes I + \ket{1}\!\!\bra{1}\otimes Z$), and single-qubit $Z$ rotations by $\theta$,  $Z_{\theta}=e^{-i\theta Z/2}$.
We will make use of the Clifford phase gate  $P\coloneqq Z_{\pi/2}$. 
Here we use ``$\circ$" to denote channel composition and ``$\comp$" for  right-to-left sequential composition of channels, i.e., $\comp_{i=1}^{n} \mathcal{E}_{i}(\rho)\coloneqq \mathcal{E}_{n}\circ\dots\circ\mathcal{E}_{1}(\rho)$.

Quantum states $\rho_{1}$, $\rho_{2}$ are commonly compared by their \textit{fidelity} $F$, given by
\begin{equation}
F(\rho_{1}, \rho_{2})=\Bigl( \tr{\sqrt{\sqrt{\rho_{1}}\rho_{2}\sqrt{\rho_{1}}}}\Bigr)^{2}.
\end{equation}
This definition also allows for comparisons between two quantum gates $\mathcal{E}_{1}$, $\mathcal{E}_{2}$.  The \emph{gate fidelity} between these two gates is defined to be
\begin{equation}
F(\mathcal{E}_1,\mathcal{E}_2)=\int \mathrm{d}\psi \, F(\mathcal{E}_1(\psi),\mathcal{E}_2(\psi)) ,
\end{equation}
where the integral is over the set of all pure states with respect to the uniform measure d$\psi$.

For a noisy implementation $\tilde{\mathcal{U}}$ of an ideal unitary gate $\mathcal{U}$, the gate fidelity $F(\tilde{\mathcal{U}},\mathcal{U})$ gives a measure of (one minus) the average case error rate of the gate.  
While the gate fidelity is a measure of the average case error, in many applications---such as computing thresholds for fault tolerance---the \emph{worst case error} is the relevant figure of merit~\cite{Aliferis2005} (quantified, for example, by the diamond norm distance between the ideal and noisy gates).  
The gate fidelity can be used to bound the worst case error rate~\cite{Wallman2015, Beigi2011,Kueng2015}.

Let $\gateset = \{\uC_r,r=1,2,\ldots,|\gateset|\}$ be a set of ideal (unitary) gates.  For each $\uC_r \in \gateset$, let $\cC_r$ be the ideal unitary gate as a superoperator and $\tilde{\cC}_r$ be a noisy approximation to this gate.       
The \textit{average gate fidelity} for the gate set $\gateset$, denoted $\bar{F}_{\gateset}$, is defined to be
\begin{equation}
\bar{F}_\gateset = \frac{1}{|\gateset|} \sum_{r =1}^{|\gateset|} F(\tilde{\cC}_r,\cC_r) \,.
\end{equation}

The RB protocol allows us to characterize the experimental implementation of a gate set $\gateset$ by estimating the value of $\bar{F}_{\gateset}$, provided that $\gateset$ forms a 
\emph{unitary 2-design}: 

\textbf{Definition (2-design):} A set of unitary gates $\gateset=\{\uC_{r}\}_r$ is a unitary 2-design if, for any quantum channel $\mathcal{E}$, the action of the twirl of $\mathcal{E}$ over $\gateset$ on an arbitrary state $\rho$ is equivalent to that of the twirl over the entire $n$-qubit unitary group~\cite{Dankert2009, Gross2007b},
\begin{align}
\frac{1}{|\gateset|} \sum_{r =1}^{|\gateset|} \cC_{r}^{\dagger} \circ \mathcal{E} \circ \cC_{r}(\rho)
 = \int \mathrm{d}U \, \mathcal{U}^{\dagger} \circ \mathcal{E} \circ \mathcal{U}(\rho) ,
\end{align}
where d$U$ is the uniform (Haar) measure.  

 For $n$ qubits, a commonly used 2-design is the $n$-qubit Clifford group~\cite{Kueng2015a, Webb2015, Zhu2015}.

\subsection{Randomized benchmarking}
\label{sec:RB}

We now briefly review RB together with a derivation (originally due to Magesan \emph{et al.}~\cite{Magesan2012}) of how RB yields an estimate of the average gate fidelity.  
In our review of this derivation, we relax the condition that the 2-design have a group structure.  
This relaxation will be important when we consider RB in the MBQC case, which will make use of non-Clifford 2-designs.

The standard RB protocol proceeds as follows.  Choose a set of unitary gates $\gateset$ that forms a unitary 2-design, and for which the inverse element of any sequence of gates can be efficiently computed.
Choose a sequence length $s$ and a number $K_{s}$ of gate sequences for that length. 
Draw $K_s$ many sequences of $s$ gates from $\gateset$ uniformly at random. 
For the $i^{\text{th}}$ sequence, $1\leq i \leq K_{s}$, denote the $j^{\text{th}}$ element of the sequence by $\uC_{j}^{(i)}$, where $1\leq j \leq s$. 
Note that each $\uC_{j}^{(i)}$ is an element $\uC_{r}\in\gateset$ from the gate set. 
For each sequence, compute $\uC^{(i)}_{s+1}\coloneqq(\uC^{(i)}_{s}\uC^{(i)}_{s-1}\dotsm \uC^{(i)}_{1})^{\dagger}$. Note that when $\gateset$ does not form a group, $\uC^{(i)}_{s+1}\notin \gateset$ in general. 

Although the sequences are ideally described by noiseless unitary gates $\uC_r$ sampled from $\gateset$, in practice these gates will be noisy.   
The noisy gates $\tilde{\cC}_j^{(i)}$ can be decomposed into a composition of the ideal unitary gate $\uC_j^{(i)}$ followed by an arbitrary CPTP map $\cD_j^{(i)}$, i.e., the noisy gate is described by 
\begin{align}
\tilde{\cC}_j^{(i)}(\rho) = \cD_j^{(i)} \circ \cC_j^{(i)}(\rho)  . \label{eq:noisegatedecomp}
\end{align}
Let $\tilde{\psi}$ denote the mixed state describing the noisy preparation of the ideal state  $\psi\coloneqq\ket{\psi}\!\!\bra{\psi}$.
The total noisy evolution of this state under the $i^{\rm th}$ sequence is then
\begin{align}
\tilde{\cC}^{(i)}(\tilde{\psi})
 \coloneqq \comp^{s+1}_{j=1} \tilde{\cC}^{(i)}_{j} (\tilde{\psi})
 = \comp^{s+1}_{j=1} \bigl[ \cD^{(i)}_j \circ \cC^{(i)}_j \bigr] ( \tilde{\psi}) .
\end{align}
At the conclusion of the sequence, a measurement described by the effects $\{\tilde{E}_{\psi}, \Id-\tilde{E}_{\psi}\}$ is performed, which is the noisy implementation of the ideal projective measurement $\{\ket{\psi}\!\!\bra{\psi}, \Id-\ket{\psi}\!\!\bra{\psi}\}$. 
This measurement gives what is known as the \emph{survival probability} for the sequence~$i$,
\begin{align}
\tr\left[ \tilde{E}_{\psi} \tilde{\cC}^{(i)}(\tilde{\psi}) \right].
\end{align} 
Its average over all $K_s$ random sequences $\cC^{(i)}$ results in the \textit{sequence fidelity}
\begin{align}
F_{\gateset}(s,K_s)
 \coloneqq \frac{1}{K_{s}} \sum^{K_{s}}_{i=1} \tr\left[\tilde{E}_{\psi}\tilde{\cC}^{(i)}(\tilde{\psi}) \right].\label{eq:avefid}
\end{align}
This can be viewed as an estimate of the average defined by the set of \emph{all} sequences of length $s$.  
As the number of sequences $K_{s}$ increases, the sequence fidelity converges to the uniform average over all sequences,
\begin{align}
F_{\gateset}(s)
 = \frac{1}{|\gateset|^{s}} \sum_{i =1}^{|\gateset|^{s}} \tr\left[ \tilde{E}_{\psi} \tilde{\cC}^{(i)}(\tilde{\psi})  \right],
\end{align}
where there are a total of $|\gateset|^s$ sequences, and each sequence $i$ is taken with equal weight in order to satisfy the 2-design condition. 
A key feature of RB is that it scales well in both the number of qubits and the sequence length $s$. 
This is due to the fact that $F_{{\gateset}}(s,K_s)$ converges quickly to $F_{{\gateset}}(s)$ in the number of sequences measured $K_{s}$~\cite{Magesan2012, Wallman2014}.

Estimating $F_{{\gateset}}(s)$ for various sequence lengths $s$ can be used to produce an approximation to the average gate fidelity  $\bar{F}_{\gateset}$. 
The original derivation~\cite{Magesan2012} is reviewed in Appendix~\ref{appA}, but presented without the assumption that $\gateset$ is a group. 
This derivation yields an exponential decay of the sequence fidelity as a function of $s$, of the form 
\begin{align}
F_{{\gateset}}(s)
 \approx A_{0} (2 \bar{F}_\gateset -1)^{s} + B_{0}\label{eq:zeroexp}
\end{align}
where $A_{0}$ and $B_{0}$ are nuisance parameters that contain information about the noise in state preparation and measurement; see Appendix~\ref{appA}. 
Equation~(\ref{eq:zeroexp}) is known as the zeroth-order expansion of $F_{\gateset}(s)$.
By performing the RB protocol above for various $s$, we can fit the zeroth-order model to the measurement data to find $\bar{F}_\gateset$~\cite{Granade2015}. 

Key assumptions in this derivation were that the noise per gate when decomposed as in Eq.~(\ref{eq:noisegatedecomp}) is Markovian 
and that it has low dependence on which gate was being applied, as well as on time, i.e., $\cD_j^{(i)} \simeq \cD$ independent of $i,j$. 
It was shown in Ref.~\cite{Magesan2012} that in this regime, the effect of including gate-dependent perturbations to the noise $\cD$ can be neglected for the purposes of calculating the average gate fidelity. 
Note that these assumptions are sufficient but not necessary---we will impose them later in Sec.~\ref{sec:RBnoise} to establish a regime under which the zeroth-order model of RB is guaranteed to be valid.

In the case when $\gateset$ is not a group, then the final sequence inverse $U_{s+1}^{(i)}$ may not be an element of $\gateset$ (and, perhaps, instead performed by changing the measurement basis).  
To directly extend the proof by Magesan \emph{et al.}~\cite{Magesan2012} to such cases, we further assume that the noise superoperator $\cD_{s+1}^{(i)}$ corresponding to the sequence inverse (or final measurement) is independent of the choice of sequence. 

\subsection{Measurement-based quantum computation}
 \label{sec:MBQC}
We now briefly review the measurement-based model for quantum computation, with a focus on the aspects that will be used in designing an RB protocol within this model.

In the measurement-based model, the task of building a quantum computer is broken into two steps: (1) prepare a cluster state~\cite{Briegel2001} with a suitable graph structure (e.g., a linear chain for single-qubit gates, or a square lattice for universal quantum computation); (2) perform single-qubit measurements on this resource, allowing for future measurement bases to be adaptively changed conditioned on past measurement outcomes~\cite{Raussendorf2003}.

\begin{figure}
\includegraphics[width=\linewidth]{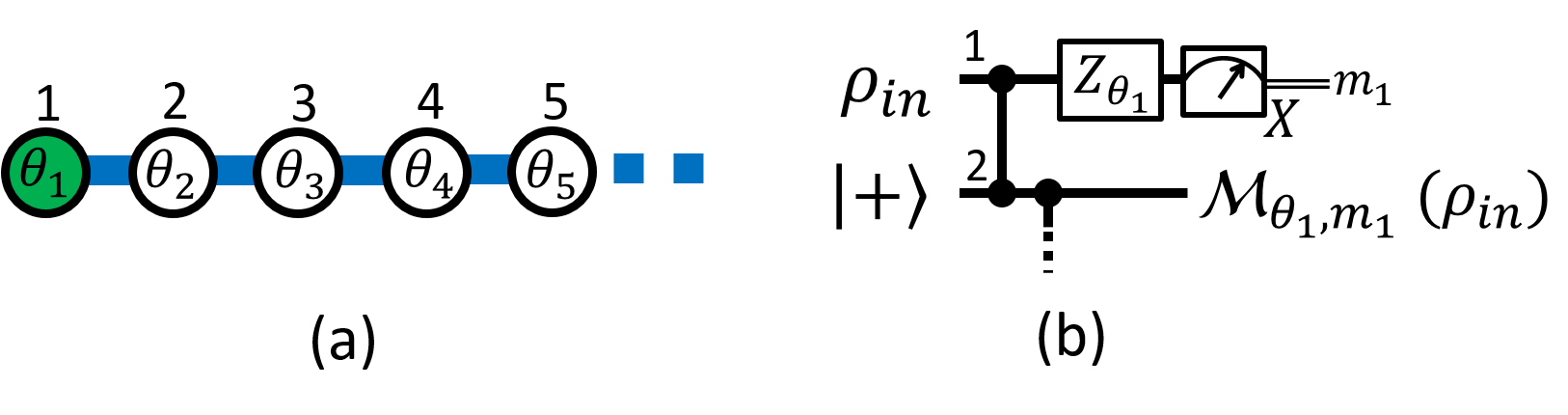}
\caption{
$\mathbf{(a)}$ Cluster state with wire graph (shown left) with an input state on the leftmost and green node. Measurement angles are labeled in the center of each node. 
$\mathbf{(b)}$ Measuring the input state in some basis in the $XY$ plane (i.e., a measurement in the eigenbasis of $X_{\theta_1} = X\cos\theta_{1}  - Y\sin\theta_{1}$) yields the output shown in the circuit on the right.}\label{fig:MBQC}
\end{figure}

For the remainder of this article, we focus our attention on linear cluster states, which allow for sequences of single-qubit gates in the MBQC model.
A linear cluster state is defined on $n$ qubits with a single-qubit input $\psi$ as   
\begin{equation}
\comp_{i=1}^{n-1} \mathcal{CZ}_{i, i+1}(\psi\otimes\ket{+}\!\!\bra{+}^{\otimes n-1})
\end{equation}
where $CZ$ gates are applied to nearest neighbors with respect the linear graph shown in Fig.~\ref{fig:MBQC}(a). 

Quantum computation proceeds via single-qubit measurements in the $XY$ plane of the qubit, as shown in Fig.~\ref{fig:MBQC}(b). 
Measuring the first cluster site in this way and obtaining the outcome $m\in\{0, 1\}$ implements the logic gate
\begin{align}
\cM_{\theta, m}(\psi)\coloneqq\cX^{m}\circ \cH\circ \cZ_{\theta}(\psi), \label{eq:MBgate}
\end{align}
where we recall that $\cX$ is the superoperator describing the unitary Pauli $X$ gate, $\cZ_{\theta}$ describes a rotation by $\theta$ about the $z$ axis, and $\cH$ is  the Hadamard gate $H$. 

In the absence of noise, both measurement outcomes are equally probable. Furthermore, the outputs only differ up to a Pauli $X$ correction, i.e., $\cM_{\theta, 1}=\cX\circ \cM_{\theta, 0}$.  

Though the gate that gets implemented after each measurement step is probabilistic (either $\mathcal{M}_{\theta_{i}, 0}$ or $\mathcal{M}_{\theta_{i}, 1}$), the overall unitary evolution due to several sequential measurements can still be made deterministic up to a known Pauli gate by using measurement feedforward, i.e.,  introducing a time ordering to the measurements and allowing the choice of the future measurement bases to depend on the outcomes of prior qubit measurements~\cite{Raussendorf2003, Nielsen2006}.   

An important exception is when $\theta_{i}$ is an integer multiple of $\pi /2\,$ $\forall i$. 
In this case, the gates are Clifford and changing the measurement angle is equivalent to flipping the measurement outcome in post processing, i.e., $\cM_{n\frac{\pi}{2}, 0}=\cM_{-n\frac{\pi}{2}, 1}$ for some integer $n$. The measurement angles do not need to be chosen adaptively based on previous measurement outcomes (as is typically required for non-Clifford gates in MBQC), and so all such measurements can be performed simultaneously.
 The final Pauli gate can be absorbed into the final measurement process. 

\section{Randomized benchmarking in MBQC}
\label{sec:RBinMBQC}

In this section,  we first give the basics of implementing RB on a linear cluster state and then we outline two schemes that use different 2-designs.  

\subsection{RB within the measurement-based model}
\label{sec:RBsetup}
For some 2-design $\gateset$, each sequence of gates $U_{r}\in\gateset$ will be implemented by measurements on a linear cluster state.  
We will analyze the use of specific gate sets in Secs.~\ref{sec:CliffRB} and \ref{sec:DeranRB}, but first we present an analysis of how RB schemes are generally performed in MBQC,  focusing on how the expected gate noise matches the noise assumptions imposed in the RB proof.  

Throughout, we assume that the same fixed number of measurements $q$ are used for all gates in $\gateset$.  
(In general, each $U_{r}\in\gateset$ may require a different number of measurements to be implemented.) 
For example, any single-qubit gate can be implemented by MBQC using $q=3$ measurements on a linear cluster state~\cite{Raussendorf2001}.  
As described by Eq.~(\ref{eq:MBgate}), the randomness of the measurement outcomes means that the logic gates performed in this way will not be deterministic and will depend on the measurement outcome.
The required total length of the linear cluster state is $(s+1)q +1$.  
(If instead the sequence inverse is incorporated into the final measurement, then only a $(s q + 1)$-long cluster state is required per run.)


\subsubsection{Noise in MBQC logic gates}
\label{sec:RBnoise}

\begin{figure*}
\includegraphics[width=0.8\linewidth]{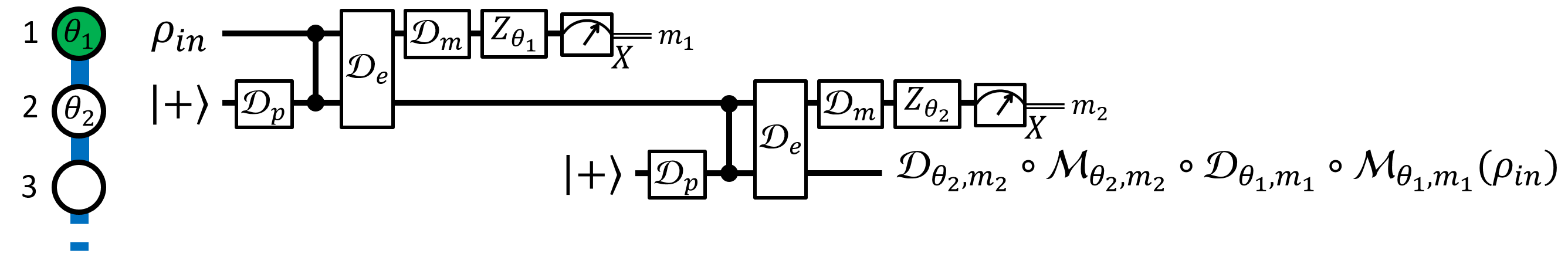}
\caption{
Here we show a small part of a noisy measurement-based quantum computation on a linear cluster state~\cite{Nielsen2006}. 
Perfect state preparation ($p$), entangling gate ($e$), and measurement ($m$) are all followed or preceded by some corresponding noise channel $\cD_{p, e, m}$. 
Each measurement step will implement $M_{\theta_{i}, m_{i}}$, along with some effective single-qubit gate error $\cD_{\theta_{i}, m_{i}}$.  
For simplicity, we have assumed that state preparation and measurement errors are the same for all cluster qubits, and therefore the effective noise channel per measurement step $\cD_{\theta, m}$ only depends on the $\theta$ and $m$. 
Note that such an error model is a generalization of those considered in Refs.~\cite{Raussendorf2003a, Lindner2009}.
}\label{fig:MBQCerror}
\end{figure*}

Noisy cluster state preparation, storage, and measurement will translate into an effective noise channel per gate as the measurement-based computation proceeds. 
The RB noise assumptions require that the errors on the cluster state be local so that gate noise from measuring different cluster qubits is uncorrelated. 
When the noise is modeled as in the circuit shown in Fig.~\ref{fig:MBQCerror},  Markovian noise in state preparations, entangling gates and measurements results in an effective Markovian noise channel per gate. 

Now consider decomposing the noisy logical gate $\tilde{\cC}_{j}^{(i)}$ as a sequence of $q$ measurements followed by a noise map, as 
\begin{align}
\tilde{\cC}_{j}^{(i)}(\boldsymbol\theta, \mathbf{m})=&\cD_{\text{seq(q)}}\circ \comp^{q}_{k=1} \cM_{\theta_{k}, m_{k}}\label{eq:multimeasure} 
\end{align}
where $\cD_{\text{seq(q)}}$ is some total noise channel after an ideal gate $\comp^{q}_{k=1} \cM_{\theta_{k}, m_{k}}=\cC_{j}^{(i)}(\boldsymbol\theta, \mathbf{m})$, and we include a dependence on $\boldsymbol\theta = ( \theta_{1}, \dots \theta_{q})$ and $\mathbf{m}=(m_{1}, \dots m_{q})$. 
The noise assumptions also require that $\cD_{\text{seq(q)}}$ be independent of time and the gate being implemented. The validity of these assumptions will depend on the relevant noise sources for
cluster state preparation and measurement.


\subsubsection{SPAM errors}

As mentioned above, errors that occur in the preparation of the cluster state can lead to logical errors in the MBQC logic gates.  
In addition to these gate errors, MBQC will also have logical state preparation and measurement (SPAM) errors.  
While the logic gates in MBQC can be robustly protected from many forms of errors by symmetry arguments~\cite{BBMR2010,Miyake2010,Else2012a,Else2012b}, this is not generally true for SPAM errors and so these can be expected to dominate in MBQC as they do in many other implementations of quantum computing.  
Nonetheless, for the purposes of RB, a natural choice of input state is $\psi=\ket{+}\!\!\bra{+}$, which is automatically ``encoded'' on the edge of the linear cluster state when prepared as in Fig.~\ref{fig:MBQC}.  After the inverse operation $U_{s+1}^{(i)}$ is applied, the final measurement is in the $X$ basis.

\subsection{RB using the Clifford group}
\label{sec:CliffRB}
Here we discuss the first of our protocols for measurement-based RB, referred to as \emph{Clifford RB}. 
The distinguishing feature of this scheme is that it uses the single-qubit Clifford group $\Clifford_{1}$ as the set $\gateset$ of logic gates.  
The Clifford group forms a unitary 2-design.

\begin{figure}
\includegraphics[width=0.8\linewidth]{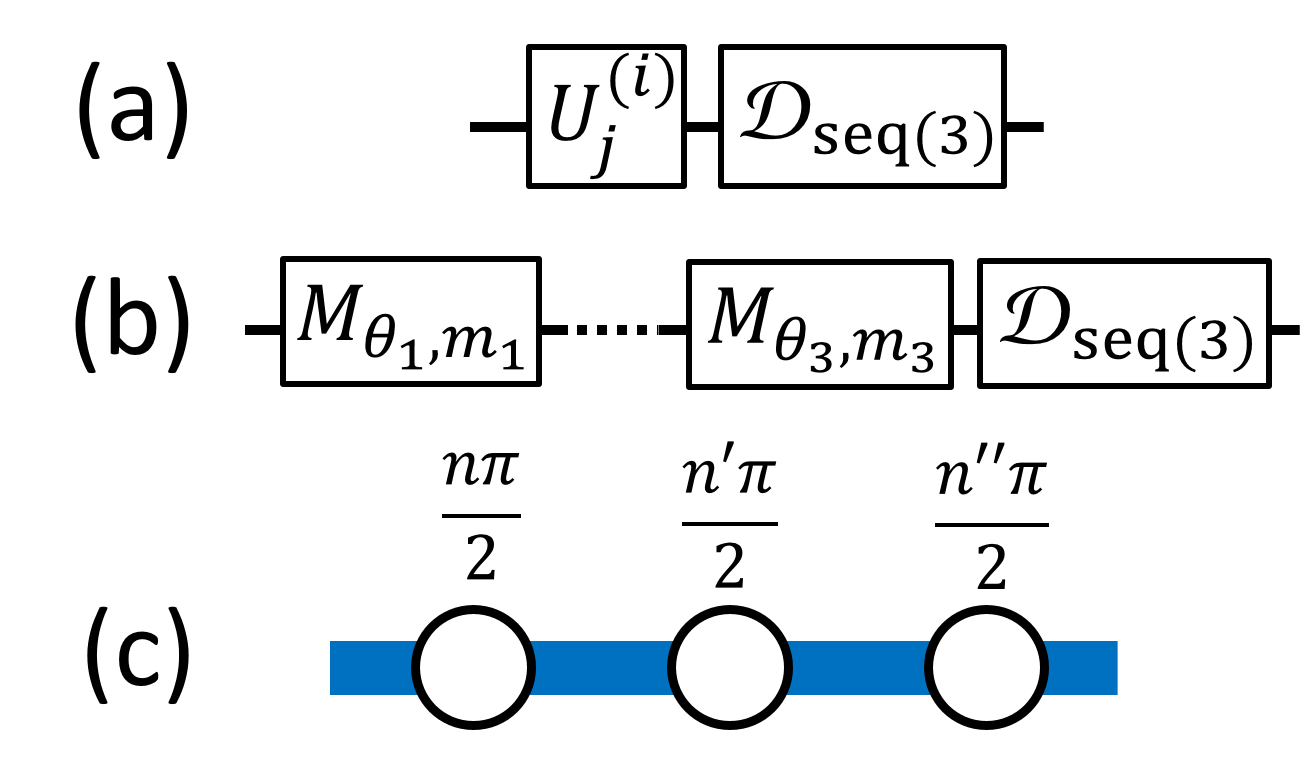}
\caption{
Each element of the 2-design (a) is implemented by making three measurements on the cluster wire [(b) and (c)]. 
We require a random sequence of Cliffords in each implementation. 
The measurement angles are all integer multiples of $\frac{\pi}{2}$ ($n, n', n'' \in \{ 0, 1, 2, 3 \})$. 
The noise operator per 2-design element $\mathcal{D}_{\text{seq(3)}}$ describes the noise added after three measurements.}\label{fig:cliffRB}
\end{figure}

We set the number of measurements per logic gate to be $q=3$, as this is the maximum number of measurements required to implement all arbitrary single-qubit Clifford gates. 
Note that this protocol can be straightforwardly extended to any $q\geq 3$ by using more measurements per gate. 
The basic building block of our scheme is the three node cluster wire shown in \fig{cliffRB}. 

Using the Clifford group simplifies the experimental setup as all measurement devices need only to be programmed to measure in either the Pauli $X$ or $Y$ basis since the measurement angles are all integer multiples of $\frac{\pi}{2}$ (see Appendix~\ref{appB} for a gate-to-measurement conversion table). 
Furthermore, we do not need to make use of measurement feedforward as the gate implemented can only differ from the desired case (e.g., $m_{i}=0$, $\forall i$) by a known Pauli gate  as
\begin{align}
\cC_{j}^{(i)}(\boldsymbol\theta, \mathbf{m})
&=\comp^{3}_{k=1} \cM_{\frac{1}{2}\pi n_{k}, m_{k}} \nonumber \\ 
&= \cX^{b_{1}}\circ \cZ^{b_{2}} \circ \Bigl(\comp^{3}_{k=1} \cM_{\frac{1}{2}\pi n_{k}, 0} \Bigr), \label{eq:Mdef}
\end{align} 
where 
\begin{align}
b_{1}&= m_{3}+m_{2}n_{3}+ m_{1}(n_{2} n_{3} +1) \\
b_{2}&=m_{2}+ m_{1}n_{2}.
\end{align}
As a consequence, each sequence can be measured simultaneously in a single time step on a linear cluster state.

The protocol begins by generating a sequence $U^{(i)}$ of length $s$ from $\Clifford_1$ uniformly at random. 
The inverse is computed in the case where all measurement outcomes are assumed to be zero. 
The corresponding measurements are made on a length-($3s+4$) linear cluster state, with the final qubit measured in the Pauli $X$ basis. 
Repeating this process $K_{s}$ times and over different sequence lengths $s$ yields an estimate for the survival probability, from which $\bar{F}_{\Clifford_1}$ can be extracted for gates implemented via three measurements. 

From Eq.~(\ref{eq:Mdef}), each $U_{j}^{(i)}(\boldsymbol\theta, \mathbf{m})$ is only implemented up to a random Pauli gate, i.e., the actual gate implemented with outcomes $\mathbf{m}=\{m_{1}, \dots  m_{q}\}$ is $U_{j}^{(i)}(\boldsymbol\theta, \mathbf{m})\in \{ I U_{j}^{(i)}(\boldsymbol\theta, \mathbf{0}), X U_{j}^{(i)}(\boldsymbol\theta, \mathbf{0}), Y U_{j}^{(i)}(\boldsymbol\theta, \mathbf{0}), Z U_{j}^{(i)}(\boldsymbol\theta, \mathbf{0}) \}$.
So long as the angles are chosen uniformly at random from the Clifford table in Appendix~\ref{appB}, the gate implemented will also be uniformly random, irrespective of the measurement outcomes. 
Thus, the indeterminism of the logic gates does not interfere with this measurement-based RB protocol. 

Next we consider the case when the probability of getting a 0 or 1 for each measurement is equally likely. 
In particular, we show how this results in a simplification of the original Clifford RB protocol. 

\subsubsection{The role of randomness}
\label{sec:randrole}
In the above, the protocol required random sequences of Clifford gates. 
However, as a result of the indeterminacy of the measurement outcomes, each chosen sequence can result in one of $4^{s}$ possible sequences occurring.  
Thus, much of the randomness required by the above protocol is redundant. 

When each cluster state measurement yields outcome 0 or 1 with equal probability, the scheme can be simplified. 
Note that $\Clifford_{1}$ can be factored into right cosets of its Pauli subgroup (ignoring phases) $\Pauli_{1}=\{ I, X, Y, Z\}$, i.e.,  {$\Clifford_{1}= \cup_{g\in \Clifsub_1} \Pauli_{1} g $}, where $\Clifsub_1 \coloneqq\{ I, P, H, P H, H P, P H P \}$. 
As a result, a random sequence of $\Clifford_{1}$ elements can be implemented by initializing the above protocol with only a random sequence of elements of $\Clifsub_{1}$. 
The larger Clifford group ($\abs{\Clifford_{1}}=24$) is generated uniformly from $\Clifsub_{1}$ by the additional random Pauli gate provided that the measurement outcomes are themselves distributed uniformly.

In general, noise on the cluster state will mean that measurement outcomes may not occur with equal probability. 
In such cases, the full Clifford RB protocol (selecting from $\Clifford_1$ rather than $\Clifsub_1$ at random) can be used. 

Alternatively, we can restore uniformity into the measurement outcome distributions to deal with problematic noise channels.  
(The alternative measurement-based RB protocol presented in the next section requires uniformly distributed measurement outcomes.)

The basic idea is to inject additional randomness into MBQC in order to restore uniformity in outcomes. 
At each measurement step $k$, we introduce a uniformly random binary variable $c_k$. When $c_k=1$, the measurement outcome is flipped,  i.e.,  $m_{k}\mapsto m_{k} + 1$ mod 2, and otherwise it is left alone. 
We have effectively defined new measurement outcome variables $m_{k}^{\prime}\coloneqq c_{k} + m_{k}$ mod 2.  
This is equivalent to applying a perfect Pauli $Z$ on the cluster qubit $k$ prior to measurement or, alternatively, locally swapping the definitions of the positive and negative $X$ axes. 
The effective measurement variable $m^{\prime}_{k}$ is a uniformly random binary variable. 
In order to use this for MBQC, the feedforward procedure must be adjusted accordingly.  
We also note that this trick will modify the effective noise channel.

Basing this scheme on the Clifford group should allow for generalization to the multi-qubit setting while preserving the advantages discussed above. 
For instance, with a universal cluster state [say, on a two-dimensional (2D) square lattice], Clifford circuits can still be implemented in a single time step as there is no need for active feedforward. 
When the measurement outcomes are uniformly distributed, random elements of the $n$-qubit Clifford group $\Clifford_{n}$ can be generated by implementing a random element of $\Clifsub_{n}$---a set containing one element from each coset of the $n$-qubit Pauli group $\Pauli_{n}$ in $\Clifford_{n}$. 
As in the single-qubit case, each element of $\Clifsub_{n}$ will be implemented along with a random Pauli, generating the full Clifford group if the measurement outcomes are uniformly random. 
Also, the inverse element of an $n$-qubit RB scheme can always be efficiently computed as a consequence of the Gottesman-Knill theorem~\cite{Gottesman1998}. 

\subsection{RB using derandomized 2-designs}
\label{sec:DeranRB}
As we saw in the previous protocol, the intrinsic randomness of MBQC can be leveraged to simplify the implementation.  
We now consider an alternative to the single-qubit Clifford group that extends this idea further: by using recently proposed measurement-based unitary 2-designs from Ref.~\cite{Turner2016}, RB can be performed using a single, fixed set of measurements and relying entirely on the measurement randomness to implement random gate sequences. 
We refer to this protocol as \emph{derandomized RB}, and it allows for the characterization of more general non-Clifford logic gates in the MBQC model.

As with Clifford RB, this scheme does not use any feedforward. 
When a linear cluster state is measured with a repeating pattern of $q$ fixed measurement bases, each set of $q$ measurements can generate up to $2^{q}$ distinct unitary evolutions. 

A necessary ingredient of this scheme is that some of the cluster qubits be measured in bases other than integer multiples of $\pi /2$. 
Otherwise, the implemented gates will only differ by a known Pauli gate [as in Eq.~(\ref{eq:Mdef})].
As $\Pauli_{1}$ is only a unitary 1-design, so too is the entire gate set, and it is insufficient for RB.

As shown in Ref.~\cite{Turner2016}, a family of 2-designs can be generated using cluster states of various lengths. 
Here we consider the simplest case: a $q=5$ sequence with measurement bases corresponding to angles
\begin{align}
\bigl(\theta_{1}, \theta_{2}, \theta_{3}, \theta_{4}, \theta_{5}\bigr) = \Bigl(\phi_{1}, \frac{\pi}{4}, \cos^{-1}\Bigl(\frac{1}{\sqrt{3}}\Bigr), \frac{\pi}{4}, \phi_{2}\Bigr). \label{eq:deranangles}
\end{align}
The resulting gate set $\gateset$ implements a unitary 2-design provided that the measurement outcomes are all equally probable. 
Note that $\phi_{1}$ and $\phi_{2}$ are free parameters, which we set equal to zero for simplicity.

In the absence of noise, the gate applied is 
\begin{align}
{\cC}_{j}^{(i)}(\mathbf{m})&=\comp^{5}_{k=1} \cM_{\theta_{k}, m_{k}} \nonumber \\ &=\comp^{5}_{k=1} \cX^{m_{k}}\circ \cH \circ \cZ_{\theta_{k}}. \label{eq:deranbasicgate}
\end{align}
Define the gate applied when all measurement outcomes are zero as 
\begin{equation}
\cQ\coloneqq \comp^{5}_{k=1} \cM_{\theta_{l}, 0}.
\end{equation}
Commuting each factor of $\cH\circ \cZ_{\theta_{k}}$ to the right in Eq.~(\ref{eq:deranbasicgate}) we get
\begin{equation}
\cC_{j}^{(i)}(\mathbf{m})=\cA^{m_{5}}_{5}\circ \cA^{m_{4}}_{4}\circ \cA^{m_{3}}_{3}\circ \cA^{m_{2}}_{2}\circ \cA^{m_{1}}_{1}\circ \cQ
\end{equation}
where 
\begin{equation}
\cA_{i}=\left(\comp^{6-i}_{k=i}\cH \circ \cZ_{\theta_{k}} \right)\circ \cZ\circ\left(\comp^{6-i}_{k=i}\cH \circ \cZ_{\theta_{k}}\right)^{\dagger}.
\end{equation}
Note that each $\cA_{i}$ is a $\pi$ rotation about some axis. 
These are expressed as $2\times 2$ matrices in Appendix~\ref{appC}. 
Therefore, the structure of each $\gateset$ element is a fixed unitary $\cQ$, followed by a sequence of $\pi$ flips, which (by construction~\cite{Turner2016}) must each be applied with probability $1/2$.
If noise in the state preparation or measurement results in a non-uniform probability distribution of measurement outcomes, then a strategy such as the one detailed in Sec.~\ref{sec:randrole} should be used to restore uniformity. 

\begin{figure}
\includegraphics[width=\linewidth]{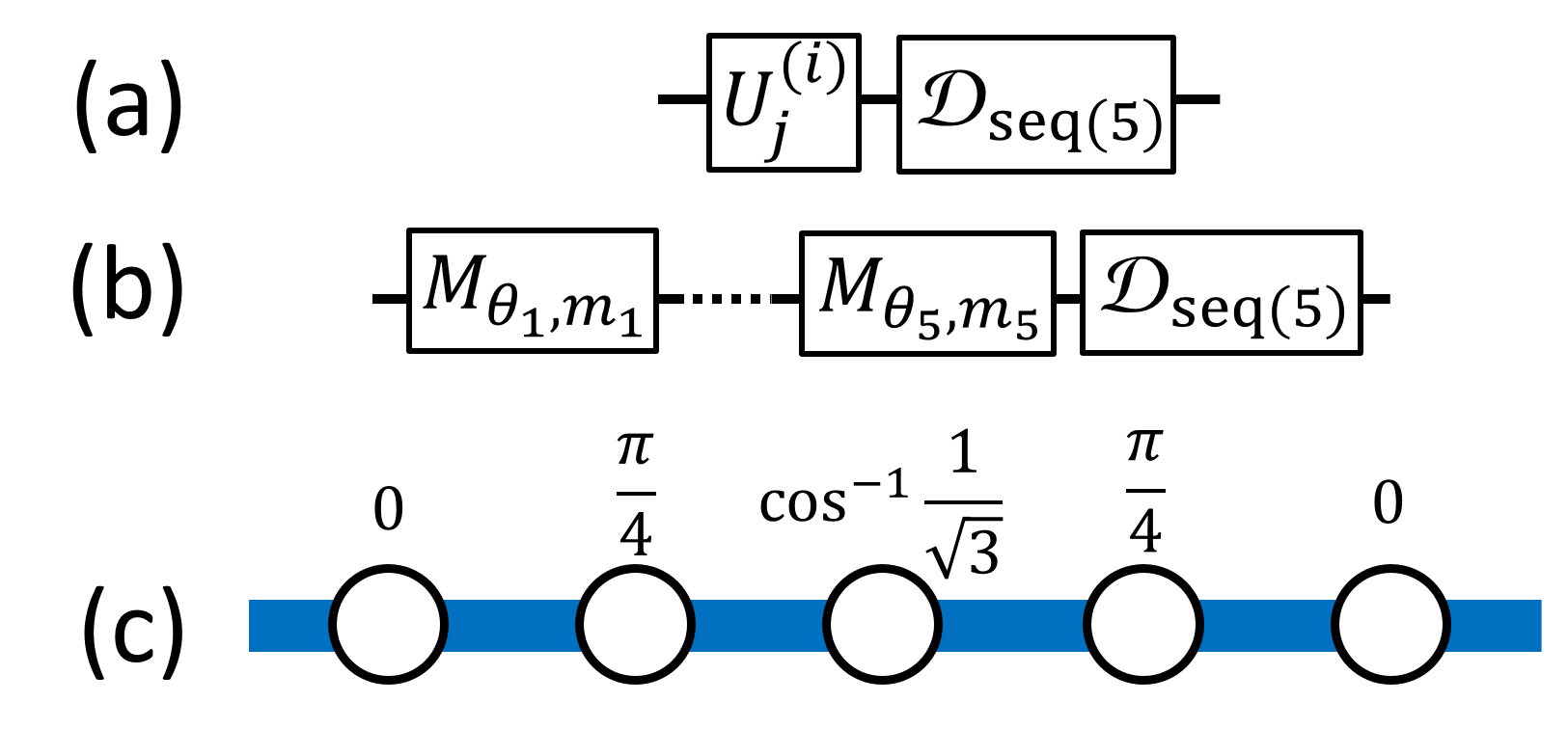}
\caption{
Each element of the unitary 2-design is implemented by making five fixed measurements on the cluster wire. 
The noise operator per 2-design element $\cD_{\text{seq(5)}}$ describes the noise added after five measurements. }\label{fig:deranRB}
\end{figure}

To use this unitary 2-design to implement a sequence of $s$ elements for RB, the sequence of measurements in Eq.~(\ref{eq:deranangles}) is repeated $s$ times on a length-($5s+1$) linear cluster state.  
The basic idea of this RB scheme is shown in Fig.~\ref{fig:deranRB}. 
We will assume that the inverse is applied via a rotated qubit measurement on the last cluster qubit. 

In this scheme, the sequence of random gates is generated by the indeterminacy of the measurement outcomes.  
As a result, the inverse element is not known \textit{a priori}. 
To determine the sequence inverse, the input state's evolution must be dynamically tracked. 
It is well known that the evolution of the state of a single qubit can be efficiently simulated classically~\cite{Nielsen2006}. 

A key advantage of this scheme's use for RB is that it uses a fixed repeating pattern of measurement angles. 
This could simplify experimental implementations, as the setup would not have to be substantially changed between different sequences.
This could also reduce noise introduced by the control in cases where sequences of gates are actively randomized.
We note that some randomness may still need to be injected to restore uniformity in the measurement outcomes as a result of noise, as discussed in Sec.~\ref{sec:randrole}. 

\section{Conclusion}
\label{sec:conclusion}
As we have shown, the basic machinery of randomized benchmarking can be translated into measurement-based quantum computation. 
Rather than interfering with the randomized benchmarking protocol, the intrinsic randomness of measurement-based quantum computation can be used to simplify it by partially (as in Clifford RB) or completely (as in derandomized RB) eliminating the need for drawing a random sequence of gates before each run.  
This work aims to establish a connection between advances in large-scale cluster state generation~\cite{Browne2005, Lindner2009, Zaidi2015, Gimeno-Segovia2015} and theoretical results for characterizing low-noise quantum devices.

For the benchmarking of gates beyond single-qubit operations, Clifford RB  generalizes naturally to the 2D square-lattice cluster state, on which the entire multi-qubit Clifford group can be implemented on arbitrary inputs via single-site measurements with angles that are integer multiples of $\pi /2$.  
Feedforward could still be performed entirely in post processing, and so a sequence of gates can still be implemented by performing all measurements on the cluster state simultaneously.  
It is known that derandomized measurement patterns can produce approximate $t$-designs in the multi qubit case~\cite{Turner2016}; however, the existence of exact multi qubit measurement-based designs is an open question. This work further motivates the search for such ensembles.

An important generalization of this work would be to characterize the validity of RB under more realistic noise sources.
Such an extension could potentially make use of higher-order expansions of the derivation by Magesan \emph{et al.}~\cite{Magesan2012} in order to deal with gate-dependent noise sources. 
An extension highly relevant to linear-optical implementations would be to find a way to deal with photon loss~\cite{Nielsen2004}, which is a non-Markovian (leakage) process. 
Dealing with this kind of noise is beyond the scope of our protocol, although recent theoretical developments have shown that the RB protocol can be adapted to such cases~\cite{Epstein2014, Chasseur2015}.
%

Another possible extension of this work could be to consider alternative gate sets $\gateset$ that are suitable for RB and can be conveniently implemented via MBQC. For instance, the dihedral RB protocol in Ref.~\cite{Carignan-Dugas2015} requires rotations about the $Z$ axis and bit flips ($X$). Within the measurement-based model on a linear cluster state, this can be straightforwardly implemented using two measurement steps per gate, where the gate specifies the angle on odd qubits and all even qubits are measured in the $X$ basis. We leave a more detailed analysis to future work.

Our work has also focused exclusively on cluster states as the resource for MBQC. 
Another generalization would be to develop RB schemes for alternative resource states such as the Affleck-Kennedy-Lieb-Tasaki (AKLT) state~\cite{BM,Darmawan,Resch}.

\section*{Acknowledgements}
The authors would like to thank H. Ball, S. Flammia, C. Granade, R. Harper, S. Kimmel, D. Markham, N. Menicucci, and J. Wallman for helpful discussions. 
This work is supported by the ARC via the Centre of Excellence in Engineered Quantum Systems (EQuS) Project No. CE110001013, and by EPSRC First Grant EP/N014812/1. 
The effort depicted is supported in part by the U.S. Army Research Office under Contract No. W911NF-14-1-0133.  
The content of the information does not necessarily reflect the position or the policy of the Government, and no official endorsement should be inferred.

\appendix
\section{Average gate fidelity derivation}\label{appA}
Here we show how estimating $F_{\gateset}(s)$ is related to $\bar{F}_{\gateset}$. 
In contrast to the original derivation in Ref.~\cite{Magesan2012}, we will do so without assuming $\gateset$ is a group.
 
\begin{widetext}
By definition,
\begin{equation}
F_{\gateset}(s)= \frac{1}{\abs{\gateset}^{s}}\sum_{i =1}^{\abs{\gateset}^{s}}\tr\left[ \tilde{E}_{\psi}\comp^{s+1}_{j=1}\left[\cD^{(i)}_{j} \circ \cC^{(i)}_{j}\right]  (\tilde{\psi}) \right]\label{Aeq:seqfid}
\end{equation}
First, we will assume we are working in a regime where the noise has little gate and time dependence. 
The zeroth-order approximation in RB makes the assumption that $\cD^{(i)}_{j}\approx \cD, \forall i, j$ with $j \leq s$, which is a good approximation in the limit of low gate and time dependence on the noise~\cite{Magesan2012}. 
It also requires that $\cD^{(i)}_{s+1}$ is independent of the choice of gate sequence $\cC^{(i)}$ , i.e., $\cD^{(i)}_{s+1}\approx \cD_{\text{inv}}, \forall i$. 
Note that this is automatically satisfied when $\gateset$ is a group (and therefore, closed under inverses) by extending the previous assumption to include $j=s+1$.
The sum in Eq.~(\ref{Aeq:seqfid}) is over all length $s$ sequences of gates from $\gateset$, and therefore it can be broken down into $s$ independent sums as follows
\begin{equation}
F_{\gateset}(s)= \frac{1}{\abs{\gateset}^{s}}\sum_{r_{s} =1}^{\abs{\gateset}}\dotsm\sum_{r_{1} =1}^{\abs{\gateset}}\tr\left[ \tilde{E}_{\psi} \cD_{\text{inv}} \circ\left( \comp^{s}_{i=1} \cC_{r_{i}} \right)^{\dagger}\circ\comp^{s}_{j=1}\left[\cD \circ \cC_{r_j}\right]  (\tilde{\psi}) \right].
\end{equation}
 
Next, we need to make use of the following: if we twirl a channel $\cD$ with the unitary 2-design $\gateset$, then we get a depolarizing channel $\cD_{\text{D}}(\rho)= p\rho+(1-p)\frac{1}{2}I$~\cite{Kimmel2014, Dankert2009}. 
That is:
\begin{equation}
\frac{1}{\abs{\gateset}}\sum_{r=1}^{\abs{\gateset}} \cC_{r}\circ\cD \circ \cC_{r}^{\dagger}(\rho)=\cD_{\text{D}}(\rho), \, \forall \rho.
\end{equation}
Crucially, this depolarizing channel has the same average fidelity as the original channel~\cite{Magesan2012}, i.e., $F(\cD_{\text{D}}, \cI)=F(\cD, \cI)$. 
Note also that $\cC\circ\cD_{\text{D}}(\rho)=\cD_{\text{D}}\circ\cC(\rho)$ for all unitary channels $\cC$. 
Then the sums implement independent twirls over the first $s$ noise channels $\cD$:
\begin{align}
& \frac{1}{\abs{\gateset}^{s}}\sum_{r_{s} =1}^{\abs{\gateset}}\dotsm\sum_{r_{1} =1}^{\abs{\gateset}}\cD_{\text{inv}} \circ\left( \comp^{s}_{i=1} \cC_{r_{i}} \right)^{\dagger}\circ\comp^{s}_{j=1}\left[\cD \circ \cC_{r_j}\right]  (\tilde{\psi}) \label{Aeq:twirl1} \\ &= \frac{1}{\abs{\gateset}^{s-1}}\sum_{r_{s-1} =1}^{\abs{\gateset}}\dotsm\sum_{r_{1} =1}^{\abs{\gateset}}\cD_{\text{inv}} \circ\left( \comp^{s-1}_{i=1} \cC_{r_{i}} \right)^{\dagger}\circ \cD_{\text{D}} \circ\comp^{s-1}_{j=1}\left[\cD \circ \cC_{r_j}\right]  (\tilde{\psi}) \label{Aeq:twirl2} \\ &= \frac{1}{\abs{\gateset}^{s-1}}\sum_{r_{s-1} =1}^{\abs{\gateset}}\dotsm\sum_{r_{1} =1}^{\abs{\gateset}}\cD_{\text{inv}}\circ \cD_{\text{D}} \circ\left( \comp^{s-1}_{i=1} \cC_{r_{i}} \right)^{\dagger}\circ \comp^{s-1}_{j=1}\left[\cD \circ \cC_{r_j}\right]  (\tilde{\psi})
\label{Aeq:twirl3} \\ &\dots = \cD_{\text{inv}}\circ\left( \comp^{s}_{i=1} \cD_{\text{D}}\right)  (\tilde{\psi})\label{Aeq:twirl4}
\end{align}
where Eq.~(\ref{Aeq:twirl4}) results from repeatedly twirling a $\cD$ operator and commuting the resulting $\cD_{\text{D}}$ leftwards as in Eqs.~(\ref{Aeq:twirl1})-(\ref{Aeq:twirl3}).
\end{widetext}
Then  
\begin{align}
F_{\gateset}(s)=&\tr\left[ \tilde{E}_{\psi} \cD_{\text{inv}}\circ\left( \comp^{s}_{i=1} \cD_{\text{D}}\right)  (\tilde{\psi})\right] \\
=&\tr\left[ \tilde{E}_{\psi} \cD_{\text{inv}}(\tilde{\psi})  \right]p^{s}+(1-p^{s})\tr\left[ \tilde{E}_{\psi} \cD_{\text{inv}} (I/2)  \right]
\end{align}
where we get $s$ copies of $\cD_{D}$ in the first line. 
Setting $A_{0}:=\tr\left[ \tilde{E}_{\psi} \cD_{\text{inv}}( \tilde{\psi} - I/2) \right]$ and $B_{0}:=\tr\left[ \tilde{E}_{\psi} \cD_{\text{inv}}(I/2)\right]$, we get 
\begin{align}
F_{\gateset}(s) \approx A_{0} p^{s} + B_{0}\label{eq:zeroexpA}
\end{align}

This is known as the zeroth-order expansion of $F_{\gateset}(s)$~\cite{Magesan2012}. 
The terms $A_{0}$ and $B_{0}$ are nuisance parameters that contain information about the noise in state preparation and measurement. 
By performing the RB protocol above for various $s$, we can fit the zeroth-order model to the measurement data to find $p$~\cite{Granade2015}. 
Then, the average fidelity of the depolarizing channel, and hence $\cD$, is simply given by $\tfrac{1}{2}(1+p)$~\cite{Magesan2012}. 

Therefore,
\begin{align}
F_{\gateset}(s) \approx A_{0} (2F_{\gateset}(s) -1)^{s} + B_{0}.\label{eq:zeroexpA}
\end{align}

\section{Clifford angles}\label{appB}
Here we provide a list of measurement angles that implement elements of the single-qubit Clifford group, assuming that all measurement outcomes are zero. 
Note that each element can be written as a product of generators $P=\begin{pmatrix}1&0\\0&i\end{pmatrix}$ and $H=\frac{1}{\sqrt{2}}\begin{pmatrix}1&1\\1&-1\end{pmatrix}$. 

To implement the full list of elements, the required measurement angles are as follows:
\begin{center}
\vspace{-0.1cm}
 \begin{tabular}{||c | c c c||}  \hline 
Gate & $ \theta_{1} $ & $\theta_{2} $ & $ \theta_{3}$ \\
  \hline \hline
$I$ & $\frac{\pi}{2}$ & $  \frac{\pi}{2} $ & $  \frac{\pi}{2} $\\
 \hline
$P$ & $0$ & $  \frac{3\pi}{2} $ & $  \frac{3\pi}{2} $ \\
 \hline
$P^{2}$ & $\frac{\pi}{2} $ & $  \frac{3\pi}{2} $ & $  \frac{3\pi}{2} $\\
 \hline
$P^{3}$ & $ 0 $ & $  \frac{\pi}{2} $ & $  \frac{\pi}{2}$\\
 \hline
$H$ & $0 $ & $  0 $ & $  0$\\
 \hline
$PH$ & $ 0 $ & $  \frac{\pi}{2} $ & $  0 $\\
 \hline
$P^{2}H$ & $0 $ & $  \pi $ & $  0 $\\
 \hline
$ P^{3}H $& $0 $ & $  \frac{3\pi}{2} $ & $  0$\\
 \hline
$HP$ & $0 $ & $ 0 $ & $  \frac{\pi}{2}$\\
 \hline
$PHP $ & $\frac{\pi}{2} $ & $ \frac{\pi}{2} $ & $  0 $\\
 \hline
$P^{2}HP$ & $ 0 $ & $  \pi $ & $  \frac{3\pi}{2}$\\
 \hline
$P^{3}HP $ & $\frac{\pi}{2} $ & $  \frac{3\pi}{2} $ & $ 0$ \\ [1ex] 
 \hline
\end{tabular}
\begin{tabular}{||c | c c c||}  \hline 
Gate & $ \theta_{1} $ & $\theta_{2} $ & $ \theta_{3}$ \\
  \hline \hline
$HP^{2}$ & $ \pi $ & $  0 $ & $  0 $\\
 \hline
$PHP^{2}$ & $ 0 $ & $  \frac{3\pi}{2} $ & $  \pi $\\
 \hline
$P^{2}HP^{2} $ & $ 0 $ & $  \pi $ & $  \pi $\\
 \hline
$P^{3}HP^{2}$ & $ 0 $ & $  \frac{\pi}{2} $ & $  \pi $\\
 \hline
$HP^{3}$ & $ 0 $ & $  0 $ & $  \frac{3\pi}{2}$\\
 \hline
$PHP^{3} $ & $ \frac{\pi}{2} $ & $  \frac{3\pi}{2} $ & $  \pi $\\
 \hline
$P^{2}HP^{3}$ & $ 0 $ & $  \pi $ & $  \frac{\pi}{2}$\\
 \hline
$P^{3}HP^{3} $ & $  \frac{\pi}{2} $ & $  \frac{\pi}{2} $ & $  \pi $\\
 \hline
$HP^{2}H $ & $  \frac{\pi}{2} $ & $  \frac{\pi}{2} $ & $  \frac{3\pi}{2}$\\
 \hline
$PHP^{2}H$ & $  0 $ & $  \frac{\pi}{2} $ & $  \frac{3\pi}{2} $\\
 \hline
$P^{2}HP^{2}H $ & $ \frac{\pi}{2}$ & $  \frac{3\pi}{2} $ & $  \frac{\pi}{2}$\\
 \hline
$P^{3}HP^{2}H $ & $ 0 $ & $  \frac{3\pi}{2} $ & $  \frac{\pi}{2}$\\ [1ex] 
 \hline
\end{tabular}
\end{center}

\section{2-design elements}\label{appC}
Here we give the $A$ matrices from Sec.~\ref{sec:DeranRB}. 
These offer a compact description of one of the unitary 2-designs discussed in Ref.~\cite{Turner2016},
\begin{align}
A_{1}=\begin{pmatrix}\frac{1}{\sqrt{3}} & -\frac{1}{6} (1+i)(\sqrt{3} + 3 i) \\ \frac{1}{6} (1+i)(3+i\sqrt{3}) & -\frac{1}{\sqrt{3}} \end{pmatrix}\\
A_{2}=\begin{pmatrix}\frac{1}{\sqrt{3}} & \frac{1}{\sqrt{3}} (1+i)\\ \frac{1}{\sqrt{3}} (1-i) & -\frac{1}{\sqrt{3}} \end{pmatrix}\\
A_{3}=\begin{pmatrix}0& e^{-i\frac{\pi}{4}}\\ e^{i\frac{\pi}{4}} &0 \end{pmatrix}\\
A_{4}=\begin{pmatrix}1& 0\\ 0 &-1 \end{pmatrix}=Z\\
A_{5}=\begin{pmatrix}0&1\\ 1 &0 \end{pmatrix}=X\\
Q=Z_{\frac{\pi}{4}} \circ H \circ Z_{\cos^{-1}\left(\frac{1}{\sqrt{3}}\right)}\circ H\circ Z_{\frac{\pi}{4}} \circ{H}
\end{align} 
Note that $A_{3}$ is an element of the Clifford group. 


\begin{thebibliography}{10}

\bibitem{Raussendorf2001}
R.~Raussendorf and H.~J. Briegel, ``{A one-way quantum computer}'',
\newblock Phys. Rev. Lett. {\bf 86}, 5188 (2001).

\bibitem{Briegel2001}
H.~J. Briegel and R.~Raussendorf, ``{Persistent entanglement in arrays of
  interacting particles}'',
\newblock Phys. Rev. Lett. {\bf 86}, 910 (2001).

\bibitem{Aliferis2006}
P.~Aliferis and D.~W. Leung, ``{Simple proof of fault tolerance in the
  graph-state model}'',
\newblock Phys. Rev. A {\bf 73}, 032308 (2006).

\bibitem{Dawson2006}
C.~M. Dawson, H.~L. Haselgrove, and M.~A. Nielsen, ``{Noise thresholds for
  optical quantum computers}'',
\newblock Phys. Rev. Lett. {\bf 96}, 020501 (2006).

\bibitem{Dawson2006a}
C.~M. Dawson, H.~L. Haselgrove, and M.~A. Nielsen, ``{Noise thresholds for
  optical cluster-state quantum computation}'',
\newblock Phys. Rev. A {\bf 73}, 052306 (2006).

\bibitem{Knill2001}
E.~Knill, R.~Laflamme, and G.~J. Milburn, ``{A scheme for efficient quantum
  computation with linear optics}'',
\newblock Nature {\bf 409}, 46 (2001).

\bibitem{Kok2007}
P.~Kok {\em et~al.}, ``{Linear optical quantum computing with photonic
  qubits}'',
\newblock Rev. Mod. Phys. {\bf 79}, 135 (2007).

\bibitem{Nielsen2004}
M.~A. Nielsen, ``{Optical Quantum Computation Using Cluster States}'',
\newblock Phys. Rev. Lett. {\bf 93}, 040503 (2004).

\bibitem{Chuang1997}
I.~L. Chuang, D.~W. Leung, and Y.~Yamamoto, ``{Bosonic quantum codes for
  amplitude damping}'',
\newblock Phys. Rev. A {\bf 56}, 1114 (1997).

\bibitem{Poyatos1997}
J.~F. Poyatos, J.~I. Cirac, and P.~Zoller, ``{Complete Characterization of a
  Quantum Process : The Two-Bit Quantum Gate}'',
\newblock Phys. Rev. Lett. {\bf 78}, 390 (1997).

\bibitem{Knill2008}
E.~Knill {\em et~al.}, ``{Randomized benchmarking of quantum gates}'',
\newblock Phys. Rev. A {\bf 77}, 012307 (2008).

\bibitem{Magesan2011}
E.~Magesan, J.~M. Gambetta, and J.~Emerson, ``{Scalable and robust randomized
  benchmarking of quantum processes}'',
\newblock Phys. Rev. Lett. {\bf 106}, 180504 (2011).

\bibitem{Magesan2012}
E.~Magesan, J.~M. Gambetta, and J.~Emerson, ``{Characterizing quantum gates via
  randomized benchmarking}'',
\newblock Phys. Rev. A {\bf 85}, 042311 (2012).

\bibitem{Wallman2014}
J.~J. Wallman and S.~T. Flammia, ``{Randomized benchmarking with confidence}'',
\newblock New J. Phys. {\bf 16}, 103032 (2014).

\bibitem{Epstein2014}
J.~M. Epstein, A.~W. Cross, E.~Magesan, and J.~M. Gambetta, ``{Investigating
  the limits of randomized benchmarking protocols}'',
\newblock Phys. Rev. A {\bf 89}, 062321 (2014).

\bibitem{Granade2015}
C.~Granade, C.~Ferrie, and D.~G. Cory, ``{Accelerated randomized
  benchmarking}'',
\newblock New J. Phys. {\bf 17}, 013042 (2015).

\bibitem{Fogarty2015}
M.~A. Fogarty {\em et~al.}, ``Nonexponential fidelity decay in randomized
  benchmarking with low-frequency noise'',
\newblock Phys. Rev. A {\bf 92}, 022326 (2015).

\bibitem{Ball2015}
H.~Ball, T.~M. Stace, S.~T. Flammia, and M.~J. Biercuk, ``{The effect of noise
  correlations on randomized benchmarking}'',
\newblock Phys. Rev. A {\bf 93}, 022303 (2016).

\bibitem{Chasseur2015}
T.~Chasseur and F.~K. Wilhelm, ``{Complete randomized benchmarking protocol
  accounting for leakage errors}'',
\newblock Phys. Rev. A {\bf 92}, 042333 (2015).

\bibitem{Kimmel2014}
S.~Kimmel, M.~P. da~Silva, C.~A. Ryan, B.~R. Johnson, and T.~Ohki, ``{Robust
  extraction of tomographic information via randomized benchmarking}'',
\newblock Phys. Rev. X {\bf 4}, 011050 (2014).

\bibitem{Johnson2015}
B.~R. Johnson {\em et~al.}, ``{Demonstration of robust quantum gate tomography
  via randomized benchmarking}'',
\newblock New J. Phys. {\bf 17}, 113019 (2015).

\bibitem{Turner2016}
P.~S. Turner and D.~Markham, ``{Derandomizing quantum circuits with measurement-based unitary designs}'',
\newblock Phys. Rev. Lett. {\bf 116}, 200501 (2016).

\bibitem{Aliferis2005}
P.~Aliferis, D.~Gottesman, and J.~Preskill, ``{Quantum accuracy threshold for
  concatenated distance-3 codes}'',
\newblock Quant. Inf. Comp. {\bf 6}, 97 (2005).

\bibitem{Wallman2015}
J.~J. Wallman and J.~Emerson, ``{Noise tailoring for scalable quantum
  computation via randomized compiling}'',
\newblock (2015), arXiv:1512.01098.

\bibitem{Beigi2011}
S.~Beigi and R.~K{\"{o}}nig, ``{Simplified instantaneous non-local quantum
  computation with applications to position-based cryptography}'',
\newblock New J. Phys. {\bf 13}, 093036 (2011).

\bibitem{Kueng2015}
R.~Kueng, D.~M. Long, A.~C. Doherty, and S.~T. Flammia, ``{Comparing
  Experiments to the Fault-Tolerance Threshold}'',
\newblock arXiv:1510.05653  (2015).

\bibitem{Dankert2009}
C.~Dankert, R.~Cleve, J.~Emerson, and E.~Livine, ``{Exact and approximate
  unitary 2-designs and their application to fidelity estimation}'',
\newblock Phys. Rev. A {\bf 80}, 012304 (2009).

\bibitem{Gross2007b}
D.~Gross, K.~Audenaert, and J.~Eisert, ``{Evenly distributed unitaries: On the
  structure of unitary designs}'',
\newblock J. Math. Phys. {\bf 48}, 052104 (2007).

\bibitem{Carignan-Dugas2015}
A.~Carignan-Dugas, J.~J. Wallman, and J.~Emerson, ``{Characterizing universal gate sets via dihedral benchmarking}",
\newblock Phys. Rev. A {\bf 92}, 060302(R) (2015).

\bibitem{Kueng2015a}
R.~Kueng and D.~Gross, ``{Qubit stabilizer states are complex projective
  3-designs}'',
\newblock arXiv:1510.02767  (2015).

\bibitem{Webb2015}
Z.~Webb, ``{The Clifford group forms a unitary 3-design}'',
\newblock arXiv:1510.02769  (2015).

\bibitem{Zhu2015}
H.~Zhu, ``{Multiqubit Clifford groups are unitary 3-designs}'',
\newblock arXiv:1510.02619  (2015).

\bibitem{Raussendorf2003}
R.~Raussendorf, D.~E. Browne, and H.~J. Briegel, ``{Measurement-based quantum
  computation on cluster states}'',
\newblock Phys. Rev. A {\bf 68}, 022312 (2003).

\bibitem{Nielsen2006}
M.~A. Nielsen, ``{Cluster-state quantum computation}'',
\newblock Reports Math. Phys. {\bf 57}, 147 (2006).

\bibitem{Raussendorf2003a}
R.~Raussendorf, ``{Measurement-based quantum computation on cluster states}'',
\newblock Int. J. Quantum Inf. {\bf 07}, 1053 (2009).

\bibitem{Lindner2009}
N.~H. Lindner and T.~Rudolph, ``{Proposal for Pulsed On-Demand Sources of
  Photonic Cluster State Strings}'',
\newblock Phys. Rev. Lett. {\bf 103}, 113602 (2009).

\bibitem{BBMR2010}
S.~D. Bartlett, G.~K. Brennen, A.~Miyake, and J.~M. Renes, ``Quantum
  computational renormalization in the haldane phase'',
\newblock Phys. Rev. Lett. {\bf 105}, 110502 (2010).

\bibitem{Miyake2010}
A.~Miyake, ``Quantum computation on the edge of a symmetry-protected
  topological order'',
\newblock Phys. Rev. Lett. {\bf 105}, 040501 (2010).

\bibitem{Else2012a}
D.~V. Else, I.~Schwarz, S.~D. Bartlett, and A.~C. Doherty, ``Symmetry-protected
  phases for measurement-based quantum computation'',
\newblock Phys. Rev. Lett. {\bf 108}, 240505 (2012).

\bibitem{Else2012b}
D.~V. Else, S.~D. Bartlett, and A.~C. Doherty, ``Symmetry protection of
  measurement-based quantum computation in ground states'',
\newblock New Journal of Physics {\bf 14}, 113016 (2012).

\bibitem{Gottesman1998}
D.~Gottesman, ``{The Heisenberg Representation of Quantum Computers}'',
\newblock (1998), arXiv:quant-ph/9807006.

\bibitem{Browne2005}
D.~E. Browne and T.~Rudolph, ``{Resource-Efficient Linear Optical Quantum
  Computation}'',
\newblock Phys. Rev. Lett. {\bf 95}, 010501 (2005).

\bibitem{Zaidi2015}
H.~A. Zaidi, C.~Dawson, P.~{van Loock}, and T.~Rudolph, ``{Near-deterministic
  creation of universal cluster states with probabilistic Bell measurements and
  three-qubit resource states}'',
\newblock Phys. Rev. A {\bf 91}, 042301 (2015).

\bibitem{Gimeno-Segovia2015}
M.~Gimeno-Segovia, P.~Shadbolt, D.~E. Browne, and T.~Rudolph, ``{From
  Three-Photon Greenberger-Horne-Zeilinger States to Ballistic Universal
  Quantum Computation}'',
\newblock Phys. Rev. Lett. {\bf 115}, 020502 (2015).

\bibitem{BM}
G.~K. Brennen and A.~Miyake, ``Measurement-based quantum computer in the gapped ground state of a two-body Hamiltonian'',
\newblock Phys. Rev. Lett. {\bf 101}, 010502 (2008)

\bibitem{Darmawan}
A.~S. Darmawan and S.~D. Bartlett, ``Optical spin-1 chain and its use as a quantum computational wire'', 
\newblock Phys. Rev. A {\bf 82}, 012328 (2010).

\bibitem{Resch}
J.~Lavoie, R.~Kaltenbaek, B.~Zeng, S.~D. Bartlett, and K.~J. Resch,
``Optical one-way quantum computing with a simulated valence-bond solid'',
\newblock Nature Physics {\bf 6}, 850-854 (2010).

\end{thebibliography}
%

\end{document}